# Fluorescent graphene quantum dots-enhanced machine learning for the accurate detection and quantification of $Hg^{2+}$ and $Fe^{3+}$ in real water samples†

Mauricio Llaver,[*a] Santiago D. Barrionuevo,[b] Jorge M. Núñez,[cdefg] Agostina L. Chapana,[a] Rodolfo G. Wuilloud,[a] Myriam H. Aguirre[fgh] and Francisco J. Ibañez[*b]

Selective, accurate, fast detection with minimal usage of instrumentation has become paramount nowadays in the areas of environmental monitoring. Herein, we chemically modified fluorescent graphene quantum dots (GQDs) and trained a machine learning (ML) algorithm for the selective quantification of $Hg^{2+}$ and $Fe^{3+}$ ions present in real water samples. The probe is obtained *via* the electrosynthesis of CVD graphene in the presence of urea, followed by functionalization with 1-nitroso-2-naphthol (NN). The functionalization with NN moieties dramatically improves the selectivity and sensitivity of the probe toward $Hg^{2+}$ and $Fe^{3+}$, as demonstrated by LODs of 0.001 and 0.003 mg L$^{-1}$, respectively. Simulations performed by time-dependent density functional theory (TD-DFT) reveals that the NN molecules within the GQDs are responsible for the florescence emission of the probe. The emission spectra profiles exhibited distinct characteristics between $Hg^{2+}$ and $Fe^{3+}$, enabling the ML model to precisely quantify and differentiate between both analytes present in natural and drinking waters. The ML results were further validated by measurements *via* cold vapor-atomic fluorescence spectroscopy and UV–vis spectroscopy. Our work demonstrates how chemical modification of GQDs, guided by an efficient ML model, markedly enhances sensitivity and selectivity in detecting harmful ions while critically reducing experiments and instrument handling.

**Environmental significance**

This work presents the development and application of fluorescent graphene quantum dots assisted by machine learning algorithms to selectively quantify and differentiate between $Hg^{2+}$ and $Fe^{3+}$ among multiple other ions in real water samples. This represents an important step forward in the integration of computational algorithms and user-friendly spectroscopic techniques to improve analytical aspects that require complex and expensive instrumentation. The ML results are further validated by measurements *via* cold vapor-atomic fluorescence spectroscopy and UV–vis spectroscopy. The efficiency of the ML model eliminates the necessity for extensive training values and validations, making it a reliable tool for precisely quantifying the Fe and Hg ions in real water samples.

## 1. Introduction

The detection of $Hg^{2+}$ and $Fe^{3+}$ ions, mainly the former, is highly required owing to their relative solubility in water and volatility in the atmosphere, which make them extremely harmful to humans, animals, and the environment.[1] The outstanding optical properties of graphene quantum dots (GQDs) have led to multiple applications, ranging from the design of efficient photoanodes[2] to new biological[3] and chemical[4,5] sensors used as fluorescence detectors of metal ions in real samples. Detection based on fluorescent carbon nanostructures is considered a superior alternative to other fluorescent probes, such as fluorescent proteins, organic dyes, and semiconducting quantum dots because of their photostability, biocompatibility, and flexibility for chemical modification[6] or doping.[7] This has led to important advances in the emission properties of GQD-based probes used for metal ion detection,[8,9] especially concerning improvements in sensitivity.[6,10] However, selectivity still presents major challenges limiting direct applications in real-life matrices. An important alternative to this limitation is to chemically modify carbon nanostructures for improving their selectivity. Surprisingly, 1-nitroso-2-naphthol (NN) molecules have not yet been incorporated into GQDs despite them being excellent color forming chelating agents with first row transition metals (*i.e.* Fe, Co, Cu, *etc.*).[11] For instance, these chelating molecules have been adsorbed on alumina for the efficient up-take of heavy metals present in wastewaters.[12] Similar molecules, such as 1-(2 pyridylazo)-2-naphthol, have also been incorporated into carbon paste electrodes for the selective determination of Co(II) present in human hair, pigs, and spinach samples.[13]

[a] *Laboratorio de Química Analítica para Investigación y Desarrollo (QUIANID), Facultad de Ciencias Exactas y Naturales, Universidad Nacional de Cuyo/Instituto Interdisciplinario de Ciencias Básicas (ICB), CONICET UNCUYO, Padre J. Contreras 1300, (5500) Mendoza, Argentina. E-mail: mllaver@mendoza-conicet.gob.ar*
[b] *Instituto de Investigaciones Fisicoquímicas, Teóricas y Aplicadas (INIFTA), Universidad Nacional de La Plata – CONICET, Sucursal 4 Casilla de Correo 16, (1900) La Plata, Argentina. E-mail: fjiban@inifta.unlp.edu.ar*
[c] *Resonancias Magnéticas-Centro Atómico Bariloche (CNEA, CONICET), S. C. Bariloche 8400, Río Negro, Argentina*
[d] *Instituto de Nanociencia y Nanotecnología, CNEA, CONICET, S. C. Bariloche 8400, Río Negro, Argentina*
[e] *Instituto Balseiro (UNCUYO, CNEA), Av. Bustillo 9500, S.C. de Bariloche 8400, Río Negro, Argentina*
[f] *Instituto de Nanociencia y Materiales de Aragón, CSIC-Universidad de Zaragoza, C/Pedro Cerbuna 12, 50009, Zaragoza, Spain*
[g] *Laboratorio de Microscopías Avanzadas, Universidad de Zaragoza, Mariano Esquillor s/n, 50018, Zaragoza, Spain*
[h] *Dpto. de Física de la Materia Condensada, Universidad de Zaragoza, C/Pedro Cerbuna 12, 50009, Zaragoza, Spain*

† Electronic supplementary information (ESI) available. See DOI: https://doi.org/10.1039/d3en00702b

Machine learning (ML) is a branch of artificial intelligence wherein algorithms are used to design models trained on known data, which can be then applied for the prediction of unknown target variables. ML offers robust and versatile computational tools to analyze systems with multiple variables, thereby dealing with multidimensional data in a highly efficient and rapid manner.[14,15] These models are already having a broad impact on everyday life aspects, such as traffic optimization,[16] disease prevention,[17] COVID-19 diagnosis,[18] chemistry,[19–24] biology,[25] and chemical sensors.[26–28] In the area of chemistry, ML-based approaches have been implemented for predicting molecular properties,[19] chemical reactions,[20] synthesis and polymer design,[21,22] improving quantum yields,[23] and precisely tuning the desired characteristics of nanomaterials.[24] For instance, ML-driven synthesis optimization has been applied to enhance the fluorescence of carbon dots (CDs), achieving quantum yields of 39.9%, which is well above commonly reported values.[23] Other ML-based approaches have been used in biology for predicting the fate of protein corona adsorbed on engineered nanomaterials,[25] as well as helping with the rational design of carbon dots for the precise tuning of physicochemical aspects, such as emission wavelength, fluorescence intensity, and solubility, to ultimately reach the desired characteristics of nanomaterials.[24] In the area of sensors, ML has been also applied with nanomaterials to build predictive correlation models for the determination of tetracyclines[27] and various metal ions in aqueous solution.[10,29,30] Related to detection, this tool has been combined with a dual-channel CDs sensor array for classifying among four different tetracycline analytes.[27] In the area of ion detection, ML has assisted CDs to demonstrate a 100% predictive accuracy toward the detection of proteins[29] and mycotoxins[30] using the gradient boosted trees (GBT) and XGBoost algorithms, respectively. It has been also incorporated with an array of carbon nanoparticles in order to discriminate between different heavy metal ions.[31]

In this work, we chemically modified GQDs to dramatically improve their selectivity toward $Hg^{2+}$ and $Fe^{3+}$ in the presence of multiple other ions present in real aqueous samples. This was accomplished *via* fluorescence spectroscopy, which is a much simpler and cost-effective technique compared to other conventional approaches.[32,33] As-synthesized nitrogen-modified GQDs were obtained by the electrosynthesis of CVD graphene-covered Ni foam in the presence of urea, and the resulting material was later functionalized with 1-nitroso-2-naphthol (NN) to improve the fluorescence intensity. The resulting nanomaterial was characterized by different techniques and successfully applied for the selective detection and quantification of $Hg^{2+}$ and $Fe^{3+}$ in the presence of multiple other ions. We incorporated a ML-based method trained with 191 excitation–emission matrices (EEMs) to quantify and discriminate between the already detected $Hg^{2+}$ and $Fe^{3+}$ ions. The aforementioned species were analyzed in river, dam, and tap waters, and the results were validated by the determinations of $Fe^{3+}$ and $Hg^{2+}$ *via* conventional spectrophotometric methods (*i.e.*, cold vapor-atomic fluorescence spectroscopy (CV-AFS) and UV–vis spectroscopy). This work represents an important advance in the use of functional nanographenes assisted by ML algorithms for the selective detection, accurate discrimination, and quantification of elemental species in drinking water samples.

## 2. Materials and methods

### 2.1. Materials and reagents

Ni foam with a thickness of 1.6 mm and 87% porosity was purchased from MTI Corp (Richmond, CA, USA). $H_2$ (99.999%) and $CH_4$ (99.999%) gases were obtained from Linde, Argentina, while urea was from Biopack (Buenos Aires, Argentina). 1-Nitroso-2-naphthol (NN, 98%) was obtained from Sigma-Aldrich, while the acetone (99.8%) and glacial acetic acid (99.7%) were from J.T. Baker. Here, 1003 mg $L^{-1}$ $Hg^{2+}$ and 1000 mg $L^{-1}$ $Fe^{3+}$ standard solutions (Merck and J. T. Baker, respectively) were used for preparing working standards. Stock solutions for the selectivity assays were from Merck and J.T. Baker. Ethyl acetate (99.8%) was obtained from Sigma-Aldrich, while potassium thiocyanate (KSCN, 99%) was from Dalton (Mendoza, Argentina). Sodium borohydride ($NaBH_4$), from Merck, was used in the CV-AFS studies.

### 2.2. Electrochemical exfoliation of graphene in the presence of urea

Urea-modified GQDs (uGQDs) were obtained following a modified electrochemical synthesis from graphene foam as previously reported by our group.[34] Briefly, a two-electrode set-up consisting in graphene-covered Ni foam and a Cu foil was immersed in a mixture of 0.1 mol $L^{-1}$ of both NaOH and urea in ethanol and subjected to 30 V. The resulting dispersion was then centrifuged at 10 000 rpm, filtered with 0.2 μm pores, and dried off and stored at 4 °C.

### 2.3. Functionalization of uGQDs with 1-nitroso-2-naphthol (NN)

A solution containing 200 mg of uGQDs in 15 mL of water was mixed with a solution containing 200 mg of NN in 30 mL of acetone. The mixture was left to react for 24 h at room temperature under constant magnetic stirring, during which the initial brownish color of both reagents turned to an intense olive-green tone. Afterwards, the mixture was left at 60 °C to evaporate the solvents, and the remaining solid was washed 4 times with 10 mL of acetone, followed by centrifugation at 1080 rpm during 15 min to remove excess NN. Finally, the purified solid was dried at room temperature and stored at 4 °C until use.

### 2.4. Collection of real samples and their fluorescence determinations

Water samples from the La Carolina river (San Luis, Argentina) and the Potrerillos dam (Mendoza, Argentina) were collected directly in clean HDPE bottles. Samples were centrifuged to eliminate sand residues and stored at −18 °C. Prior to the analysis, aliquots were filtered through 0.45 μm pore-size PTFE membrane filters. Tap water was sampled directly in a clean beaker. In all cases, 5.0 mL of the samples was treated with 50 μL of a 2.0 mol L$^{-1}$ acetic acid/sodium acetate pH 4.0 buffer before analysis. Phosphate and TRIS buffers (0.02 mol L$^{-1}$) were used for adjusting the pH to 7.0 and 9.3, respectively, while HCl was used for adjusting the pH to 1.0. In all cases, fluorescence analyses were run either on the treated samples or 5.00 mL aliquots of buffered standard solutions. Here, 70 μL of 3.00 mg mL$^{-1}$ aqueous NN-uGQDs were then added, and the EMMs of the mixtures were immediately obtained. Measurements were carried out with $\lambda_{em}$ in the 310.0–350.0 nm range and with $\lambda_{ex}$ between 375.0–499.0 nm, with data intervals of 2.0 nm. The excitation and emission slits were fixed at 5.0 and 10 nm, respectively, and the scan rate was 600 nm min$^{-1}$.

### 2.5. Instrumentation

An RF-6000 spectrofluorophotometer (Shimadzu, Kyoto, Japan) was used for all the fluorescence measurements, with a non-fluorescent quartz cell. EEMs were obtained using the 3D built-in method included in the LabSolutions RF software (Shimadzu). Attenuated total reflectance-Fourier transform infrared spectroscopy (ATR-FT-IR) was performed using a Perkin Elmer Spectrum Two instrument (Beaconsfield, UK) with a Universal ATR module. Imaging of the nanostructures and structural characterization was performed by high-resolution transmission electron microscopy (HRTEM), using an aberration-corrected Titan3 60-300 microscope (Thermo Fisher Scientific, MA, USA) operating at 80 kV, at room temperature. X-Ray photoelectron spectroscopy (XPS) data were collected using a K-Alpha+ instrument (Thermo Fisher Scientific), with Al Kα radiation at 1486.69 eV (150 W, 10 mA), a charge neutralizer, and a delay line detector consisting of 3 multi-channel plates. The survey spectra were recorded from −5 to 1350 eV (pass energy = 150 eV, 2 sweeps), using an energy step size of 1 eV and a dwell time of 100 ms. High-resolution spectra for C 1s, O 1s, and N 1s were recorded in the appropriate regions, with a 0.1 eV energy step size. UV–vis spectroscopy was performed with a Shimadzu UV 1800 spectrophotometer. For CV-AFS, an AF-640A Rayleigh atomic fluorescence spectrometer (Rayleigh Analytical Instrument Corp., Beijing, China) was used, with a hollow-cathode Hg lamp. A Horiba F-51 pH meter (Kyoto, Japan) was used for the pH determination. Ultrapure water was obtained from a Milli-Q water purification system and used throughout the study (APEMA, Buenos Aires, Argentina).

### 2.6. Computational simulations

Calculations were carried out using the ORCA Quantum Chemistry Package.[35] The ground state geometries of the simulated systems were first optimized with density functional theory (DFT) using the BP86 functional and a polarized valence double-zeta basis set (SVP). All dangling bonds were passivated with H atoms. Additionally, all geometries were solvent-corrected with the conductor-like polarizable continuum model (CPCM) method for water solutions. Absorption and fluorescence spectra were calculated using optimized ground state geometries and the TD-DFT method with the BP86 functional and SVP basis set, plotted and analyzed using the ORCA Advance Spectral Analysis module, with the IMDHO model.[36] All the computations were carried out in parallel in a 6 core and 32 GB RAM system.

### 2.7. Data analysis

All data processing (except when noted) and model building was carried out with Python 3.9.12, using Jupyter Notebooks (Project Jupyter, https://jupyter.org). Data was pre-processed using the Pandas (https://pandas.pydata.org) and NumPy (https://numpy.org) libraries, while all ML models were built using scikit-learn (https://scikit-learn.org/stable/). All the involved codes are available as an additional ESI† (a Jupyter Notebook called code_esi.ipynb). A schematic summary of the data processing steps is presented in Fig. S1, in the ESI.† Initially, the spectrum for each combination of Fe$^{3+}$ and Hg$^{2+}$ concentrations was employed as a matrix of fluorescence intensity as a function of $\lambda_{ex}$ and $\lambda_{em}$. Each of these matrices was first flattened to a vector of intensities and normalized ($I/I_0$) to avoid inter-day fluctuations. These vectors were then stacked into a final two-dimensional (2D) array as a function of the corresponding concentrations of Fe$^{3+}$ and Hg$^{2+}$. This final array was used to generate a second 2D array, but this time, using the reciprocal intensities quotient ($I_0/I$). Then, both arrays were used directly to train and test the studied ML models. A total of 191 EEMs, including 95 duplicate measurements at different Hg$^{2+}$/Fe$^{3+}$ combinations, plus a blank, were obtained for the training and cross-validation of the models. After obtaining the data inputs, three linear models were compared as predictors; namely, classic linear, ridge, and LASSO regressions. In ridge regression, the linear coefficients are squared and constrained by a factor ($\alpha$), penalizing those predictors that influence the model the most, reducing the chances of overfitting. LASSO regression works similarly, albeit considering the coefficients directly, not only penalizing the most influential features with an alpha factor, but also eliminating those that have little or no influence at all.[37] Both classic linear and ridge regression models are preceded by a recursive feature elimination (RFE) step to reduce dimensionality,[38] while LASSO does this as part of the training process. In the cases of ridge and LASSO regressions, grid-search cross-validations are carried out for hyperparameter tuning, using $K$-fold validation. This approach divides the dataset into $K$ equal-sized groups, and uses the first group as the test set and the ($K$–1) remaining groups as the training set. This is repeated $K$ times until all the groups have been used as test sets and then reports a single average accuracy metric for all the iterations.[39] When coupled with a grid-search, this process is repeated for all of the proposed hyperparameter combinations, allowing for a full optimization and validation of the model.[40] Throughout this work, a value of $K$ = 5 was used, and $R^2$ was chosen as the accuracy metric for predictions. In the case of classical linear regression, where no hyperparameter tuning was necessary, cross-validation was carried out directly, using $K$ = 5. The output files from the experimental excitation–emission matrices (EEMs) with known $Hg^{2+}$ and $Fe^{3+}$ concentrations had to be pre-processed prior to the application of the machine learning models used for the prediction of analyte concentrations in real samples.

## 3. Results and discussion

### 3.1. Experimental set-up and characterization

Fig. 1 shows all the experimental steps involved in this work to achieve a selective and sensitive probe. The experiment involves the electrosynthesis of GQDs in the presence of urea followed by the chemical modification of 1-nitroso-2-naphthol (NN) for improving the fluorescence and selectivity, respectively, to ultimately combine the probe with a machine learning algorithm capable of predicting and quantifying $Fe^{3+}$ and $Hg^{2+}$ among multiple ions in real matrices.

Fig. 2(a) shows an HR-TEM image of the as-synthesized uGQDs showing monodisperse and ultrasmall (∼3.0 nm) mean size diameter nanocarbons as indicated by the histogram (next to the figure) and a monomodal narrow log normal distribution with an 1.8 nm full-width half-maximum (FHWM), respectively. Fig. 2(b) shows a magnification of a selected uGQD, enclosed by a red square in Fig. 2(a), in which a hexagonal pattern can be observed, indicating highly crystalline structures. The fast Fourier transform (FFT) analysis shown in Fig. 2(c) for the same sample exhibited an hexagonal pattern with diffraction points corresponding to ∼0.21 and 0.25 nm, forming angles of 55° and 70°, assigned to a slightly tilted hexagonal closed-packed configuration that may have been aroused from a few-layer graphene (FLG) AB-stacked graphene (hexagonal, $P6_3/mmc$ #194, $a = b$ = 2.46 Å, $c$ = 6.70 Å, $\alpha$ = 60°, $\beta$ = 120°).[41,42] Few-layers structures arranged in an AB-stacked fashion are generally expected from graphene grown on Ni, since previous Raman experiments performed by our group[34] and others[43] have determined the presence of rotated and AB-stacking configurations. Therefore, we could ensure that the as-synthesized GQDs retained the original structure of FLG graphene but in smaller subdomains capable of being dispersed in water for further applications. Concerning the chemical environment of the carbon nanostructures, Fig. 3 shows the deconvoluted C 1s, O 1s, and N 1s XPS spectra for the as-synthesized uGQDs and 1-nitroso-2-naphthol-modified uGQDs (NN-uGQDs). Additional XPS spectra of pristine 1-nitroso-2-naphthol (NN) are shown in S2 (ESI†). The XPS spectra depict interesting features related to the chemical functionalization of the as-synthesized uGQDs later modified with NN groups. Regarding the former, N appeared to be part

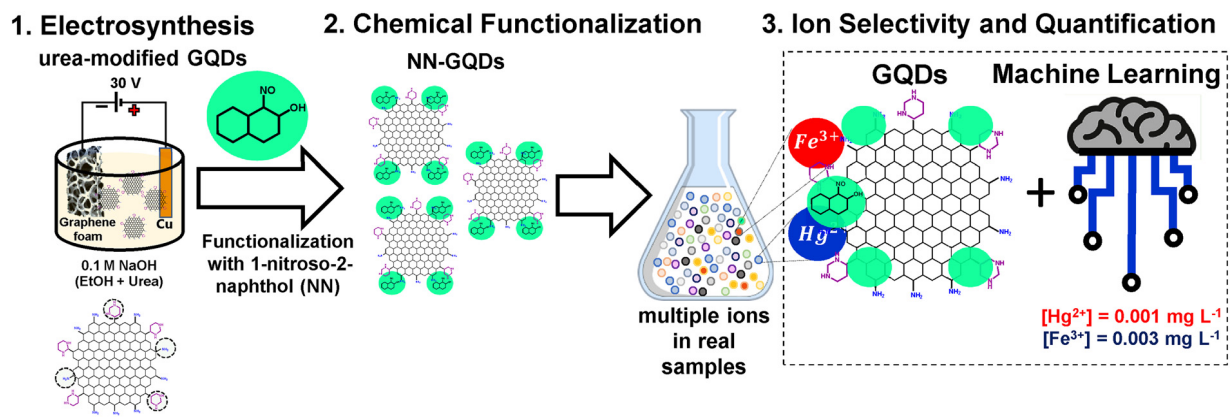

Fig. 1 Experimental scheme of all the necessary steps to built up the sensor. Step 1 indicates the electrosynthesis of graphene foam in the presence of urea to form urea-GQDs. Step 2 shows the chemical modification of urea-GQDs with the 1-nitroso-2-naphthol (NN) chelating agent for the selective and sensitive detection of $Fe^{3+}$ and $Hg^{2+}$ ions in solution. Step 3 incorporates a ML algorithm to boost the quantification of those metal ions.

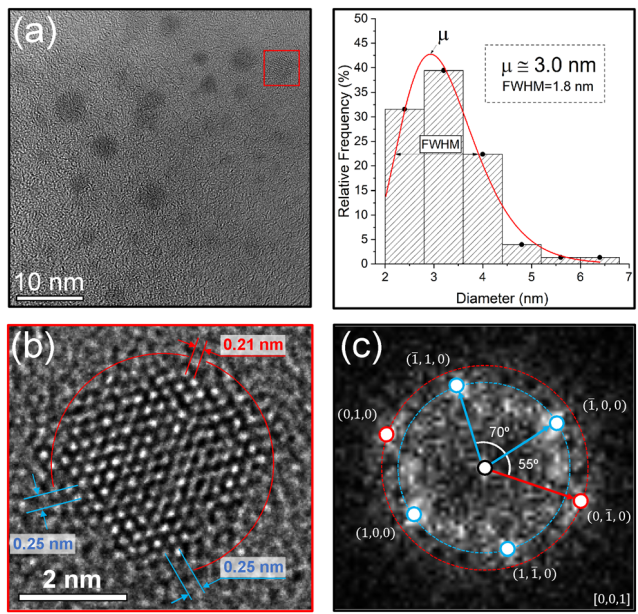

Fig. 2 (a) HR-TEM image of the as-synthesized and as-deposited nitrogenated graphene quantum dots (uGQDs), along with a histogram depicting their size distribution. (b) Magnification of a selected uGQD circumscribed by a red square in (a). (c) Fast-Fourier transform of the selected uGQD indicating tilted bi- or few-layer graphene quantum dots.

of pyridinic, amino, and nitrile groups, whose characteristic signals were also found in the corresponding ATR-FT-IR spectra (Fig. 3). In addition, the XPS C 1s, O 1s, and N 1s spectra for NN-uGQDs exhibit C–N=O moiety signals that were also part of the pristine NN spectra. This was also consistent with the ATR-FT-IR spectra of the NN-uGQDs (Fig. 3G), whose signals coincide with those observed for NN. Additionally, the XPS signals assigned to C–C, C=C, and C–O are characteristic of the carbogenic core of GQDs and match up with signals found in the ATR-FT-IR spectra at 1583, 1430, and 880 cm$^{-1}$, respectively.

Fig. 4 shows the excitation–emission matrices (EEMs) obtained for the pristine NN and urea-synthesized GQDs (uGQDs) before and after their functionalization with NN. First of all, the free molecule 1-nitroso-2-phenol (NN) did not exhibit any fluorescence, likely due to its poor solubility in water. For comparison, the EMM for the GQDs synthesized without urea (control) is shown in section S3 of the ESI.† The overall spectrum for uGQDs was similar to that for the GQDs except that the former exhibited a higher fluorescence intensity and a distinct emission when excited at 315 nm. Furthermore, Fig. 4 shows the important features in the EEMs of uGQDs before and after their functionalization with NN, discussed as follows. First, the uGQDs exhibited an excitation-independent behavior, as determined by the single band ∼375 nm, obtained at 225, 256, and 315 nm excitation wavelengths ($\lambda_{ex}$). It has been proposed that this behavior comes from a low number of surface defects and highly uniform size distribution of the product.[44] Second, the spectrum for the NN-uGQDs showed a significant emission intensity (456 nm) when excited at 326 nm, which did not occur in the case of the uGQDs. Finally, since the spectrum of pristine NN showed no emission within the studied wavelengths range, this suggests that the observed changes in the emission of uGQDs came from their chemical functionalization, rather than from an excess of complexing reagent that may have been present within the modified nanomaterial.

To gain insights into the fluorescence mechanism, we calculated the absorption and emission spectra of the NN molecules by means of TD-DFT calculations. Fig. 5 shows the results from ORCA quantum chemistry package used for simulating the excited states of the molecule and its emission and absorbance spectra. Considering the calculated and observed emission wavelengths, we propose that the NN molecule is responsible for the fluorescence of the probe, showing an emission intensity one order of magnitude higher than the isolated molecule in water. Our hypothesis considers that the energy absorbed by the carbogenic core is readily transferred to the NN molecules due to the proximity between them, ultimately leading to an enhanced emission via the NN moieties. In this sense, it is important to highlight that the functionalization with NN not only increased the water solubility, but also became a key factor for further improving the selectivity to certain ions. As a partial conclusion, significant differences were observed in the emission characteristics between the as-synthesized uGQDs and the NN-functionalized uGQDs.

### 3.2. Distinct fluorescence responses to $Hg^{2+}$ and $Fe^{3+}$

Fig. 6 shows the selected EEM spectra of NN-uGQDs aqueous dispersions before and after the addition of 0.50 mg L$^{-1}$ and 0.45 mg L$^{-1}$ of $Fe^{3+}$ and $Hg^{2+}$, respectively, excited at 326 nm wavelength ($\lambda_{ex}$). As can be observed, the emission profiles were similar regarding the drop in intensity but different with respect to the final emission wavelength. It could be observed that in the presence of $Fe^{3+}$ the emission maximum was slightly blue-shifted, while $Hg^{2+}$ did not display any change. This behavior is further confirmed in Fig. S4 in the ESI,† which displayed the same fashion even at lower concentrations of $Fe^{3+}$ and $Hg^{2+}$ ions (0.20 mg L$^{-1}$). These particular characteristics were crucial for the ML model, because such distinct differences in the EEM profiles could be readily assimilated by the algorithm for later discriminating and quantifying between both analytes. Even though the exact mechanism for selectivity and fluorescence quenching remains unclear, we believe that the latter may occur due to the proximity of the metal cation upon complex formation, which in turn, induces non-radiative recombination of the generated exciton via an electron-transfer step. This reduces the HOMO–LUMO gap of the probe, thus lowering the probability of radiative relaxation, which causes damping in the emission intensity and eventually a shift in the emission wavelength. The slight blue-shifts and no shift at all experienced upon the addition of $Fe^{3+}$ and $Hg^{2+}$, respectively, rule out the possibility of an aggregation-induced mechanism since the expected outcome would have been a significant red-shift instead.

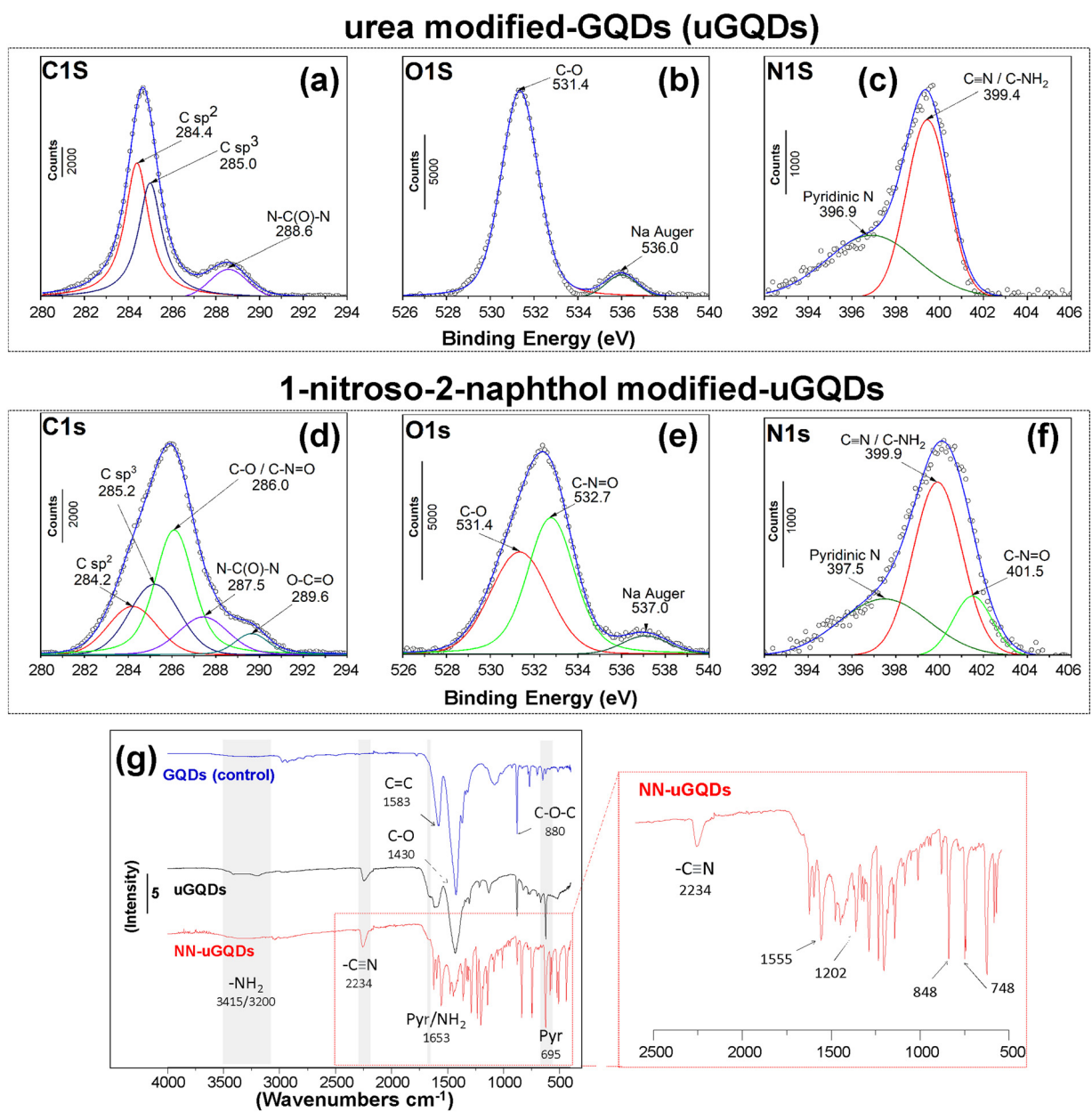

**Fig. 3** X-Ray photoelectron spectra (XPS) of C 1s, O 1s, and N 1s for the uGQDs (A, B, and C, respectively) and 1-nitroso-2-naphthol-modified uGQDs (NN-uGQDs) (D–F), including deconvolution of the instrumental signals. ATR-FT-IR spectra of GQDs (control), uGQDs, and NN-uGQDs (G). Next to panel G, there is a zoom-in of the ATR-FT-IR spectrum of NN-uGQDs. Spectra in G are off-set for better comparison.

### 3.3. Optimal pH for improving the selectivity

The selectivity of GQD-based probes is usually a factor that limits their applications, since they tend to respond to several species and molecules, thus seriously compromising their differentiation. Fig. 7 shows the responses of the uGQDs and NN-uGQDs tested at pH = 4.0, 7.0, and 9.3 upon the addition of 1.00 mg mL$^{-1}$ of selected ionic species. From the calibration curve, we obtained linear responses ranging from 0.042–1.00 mg L$^{-1}$ for both analytes. Accordingly, 0.042 mg mL$^{-1}$ was established as the optimal probe concentration for the further studies. As can be observed in Fig. 7(a), the as-synthesized uGQDs were sensitive to all the studied species, showing no selectivity regardless of the pH. On the other hand, the NN-uGQDs (Fig. 7b) were relatively more sensitive to $Hg^{2+}$ and $Fe^{3+}$ at pH 4.0, as noticed by the selective quenching observed in the presence of both species. These results were used as a starting point for a deeper study on the effect of other potential interfering species.

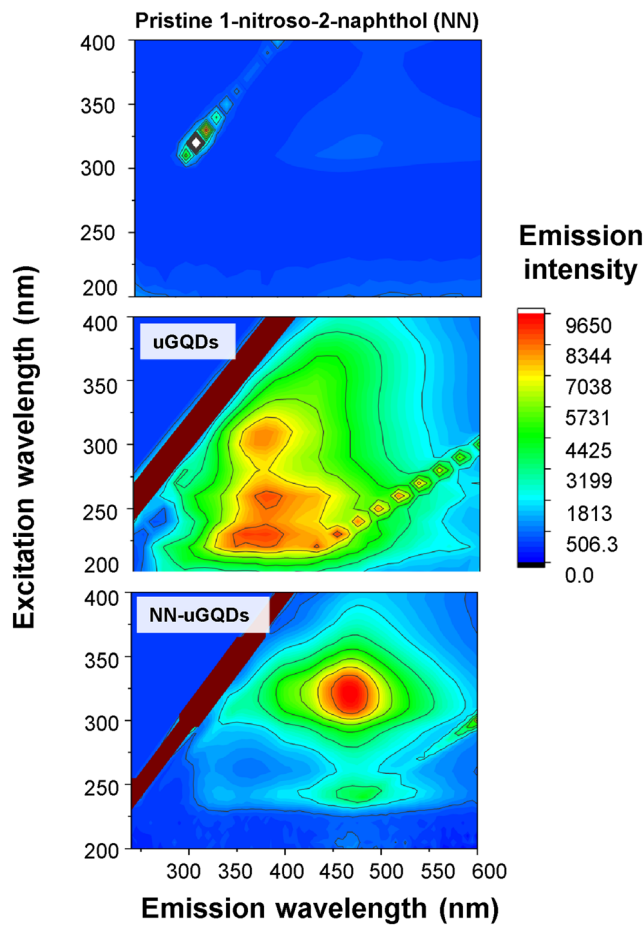

Fig. 4 Excitation–emission matrices (EEM) for pristine 1-nitroso-2-naphthol (NN), urea-modified graphene quantum dots (uGQDs), and 1-nitroso-2-naphthol-modified uGQDs (NN-uGQDs). For clarity proposes, the high-intensity emission from scattering ($\lambda_{em} = \lambda_{ex}$) has been colored brown in the middle and bottom spectra.

Fig. 8 shows the relative emission intensities for NN-uGQDs at pH 4.0, in the presence of a wide variety of cationic and anionic species run at 1.00 mg mL$^{-1}$ concentrations, which are well above those expected in real samples. The data were obtained with $\lambda_{ex}$ = 326 nm and $\lambda_{em}$ = 426 nm. The figure indicates that in the presence of several cations and anions, besides Hg$^{2+}$ and Fe$^{3+}$, the emission characteristics of the NN-uGQDs did not suffer quenching. Considering the results presented in Fig. 7 and 8, we ascribe the observed behavior to a combination of factors. First of all, it is apparent from Fig. 7(a) that the studied analytes acted as the strongest quenchers of the emission of the as-synthesized uGQDs. This could be attributed to the electronic properties of Hg$^{2+}$ and Fe$^{3+}$, which are known to reduce the fluorescence intensity of GQDs very efficiently via electron-transfer mechanisms.[45,46] At the same time, the chelating properties of NN are thought to modify the surface of the uGQDs, affecting their interaction with the ions. Furthermore, by considering the excess electron pairs in the NN structure and the ability of these metals to acquire electrons in their s and p orbitals, the formation and stability of complexes between them becomes feasible.[11,47] Therefore, these results could be explained by the premise that NN facilitates the formation of surface complexes preferentially with Hg$^{2+}$ and Fe$^{3+}$ (at pH 4) as shown in Fig. 7(b). Other ions would be at a greater distance from the core of the uGQDs, thus producing no effect on the fluorescence of the probe. However, further studies are needed to conclusively determine the mechanistic aspects of this particular interaction.

### 3.4. Stability of the probe

Concerning the optimal storage and conservation conditions, we daily tested the emission of a stock dilution of 3.00 mg mL$^{-1}$ prepared in ultrapure water for a period of 2 months kept at 4 °C with no protection against ambient light. No apparent changes in the emission characteristics were observed within this time window. To test their photostability, 0.042 mg mL$^{-1}$ solutions of NN-uGQDs were subjected to 60 min irradiation from a 15 W UVC lamp (maximum emission at 253.7 nm, according to the manufacturer). No apparent photobleaching of the probe was observed, therefore it outperformed other GQD-based probes developed by our group.[4,5] The response time of the fluorescence probe was also evaluated and involves changes in the emission at 456 nm ($\lambda_{ex}$ = 326 nm) in the presence of 1.00 mg mL$^{-1}$ Hg$^{2+}$ and Fe$^{3+}$, both simultaneously and separately. The results, in Fig. S5 of the ESI,† revealed that at least 15 min is needed after the addition of Fe$^{3+}$ to achieve a stable turn-off signal, unlike Hg$^{2+}$, which readily occurs. Importantly, when the signal is stabilized, it remained constant for at least 60 min.

### 3.5. Machine learning-based data analysis

To quantize the effect of Hg$^{2+}$ and Fe$^{3+}$ on the probe emission, various linear ML models were trained and tested using the reciprocal relative intensity data ($I_0/I$) obtained from 190 spectra. We initially considered the relative intensities ($I/I_0$) as features, but the best results in terms of accuracy were achieved when using the reciprocal intensity ($I_0/I$). Furthermore, neither min–max nor normalization standardization helped improving the results, so the data were used ipso facto. The best results being obtained using the reciprocal relative intensity is consistent with the Stern–Volmer model,[48] which proposes a linear relationship between $I_0/I$ and the quencher concentration:

$$\frac{I_0}{I} = 1 + K_{SV}[Q] \quad (1)$$

where $K_{SV}$ corresponds to the Stern–Volmer constant and [Q] to the quencher concentration. Since each EEM included 1323 data points, dimensionality reduction became necessary prior to classical linear and ridge regression (see the experimental section) in order to reduce noise and overfitting issues (known as "the curse of dimensionality").[49] Visual inspection of the EEMs showed that for $\lambda_{em} < 398$ nm, the data was highly noisy. Removing these data points resulted in spectra with 1071 data points, and greatly improved the cross-validation accuracies without compromising important information. Recursive feature elimination (RFE) was then chosen for further dimensionality reduction. In this method, the model is fitted to the entire dataset, and data points with the least impact on the model are removed. This is repeated recursively in sufficient fit/removal cycles, until the desired number of features (the objective) is reached. This is also accompanied by the tuning of hyperparameters for ridge regression, in which the parameters alpha (the strength of the regularization) and solver (the numerical approach to perform the model fit) were optimized for each RFE objective. In the case of LASSO regression, RFE was not necessary, since feature selection is part of the algorithm. In this case, the tuned hyperparameters were alpha (analogous to that of ridge regression), max_iter (the maximum number of iterations performed during the building of the model), and tolerance (the minimum accepted value for the improvement of accuracy between iterations). Bearing this in mind, initial analyses of the 191 available EEMs (totaling 204 309 data points) were carried out with the three linear models, including RFE and hyperparameter tuning when corresponding.

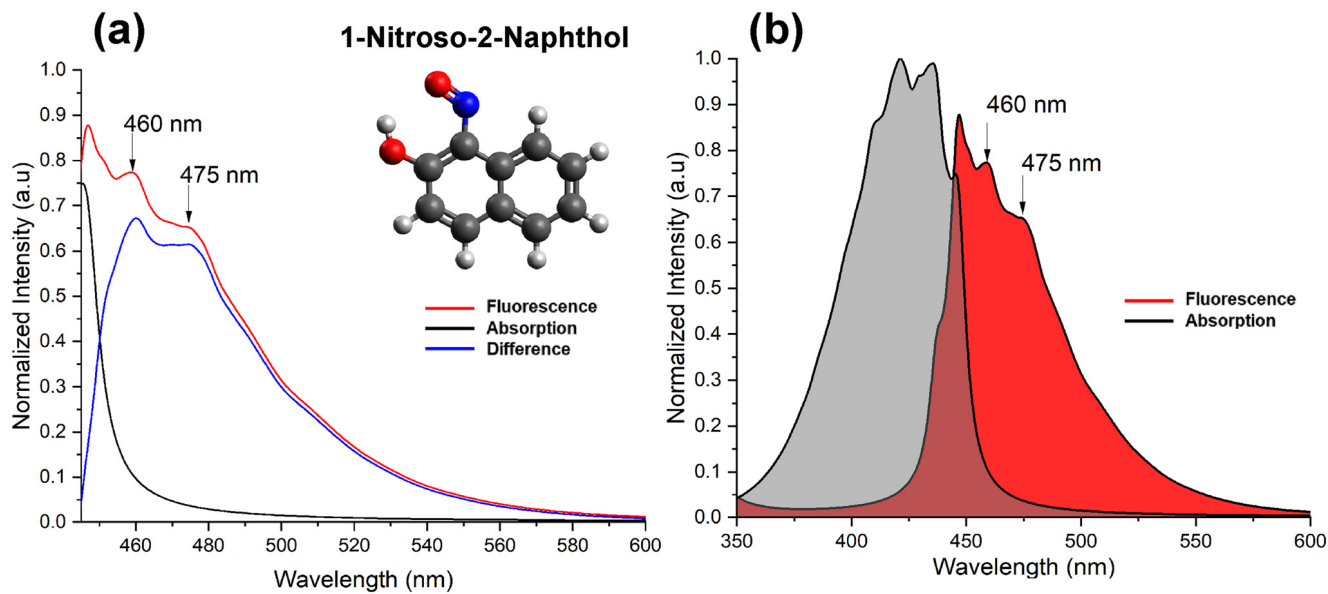

**Fig. 5** Simulated absorbance and fluorescence spectra for 1-nitroso-2-naphthol dissolved in water, indicating the most probable fluorophore emission (a). Fluorescence emission is depicted in red, absorbance in black, and the difference in blue. A representation of the molecular structure of 1-nitroso-2-naphthol is presented, with the color gray depicting C atoms, white depicting H atoms, blue depicting N atoms, and red depicting O atoms. Simulated absorbance and fluorescence are shown together colored with gray and red, respectively (b).

Fig. 9 shows the average $R^2$ values obtained for the cross-validations (with $K = 5$) as a function of the RFE objective for both, the classical linear and ridge regressions. As can be seen, excellent accuracies, with $R^2$ as high as 0.992, were achieved in both cases. However, the ridge model required more than double the number of features than the classical linear model to reach such accuracy. Taking this into account, the classical linear model was considered superior, achieving the maximum accuracy with 270 features. The figure also shows that for low objectives (that is, using a single or only a few wavelength combinations for calibration) the results were quite unacceptable, which was the original limitation that motivated the use of an ML-based approach for data analysis. Regarding the ridge model, the optimal alpha parameter was found to be 0.0001, and the optimal solver method was singular value decomposition (SVD). The LASSO model, on the other hand, yielded a single 0.977 $R^2$ value, with an optimal alpha hyperparameter of $1 \times 10^{-6}$, and $1 \times 10^{-5}$ and $1 \times 10^{-6}$ as the optimal maximum iterations and tolerance, respectively. In conclusion, taking these results as an initial assessment, a classical linear model with 270 features was considered optimal and therefore applied for the subsequent analysis of real samples.

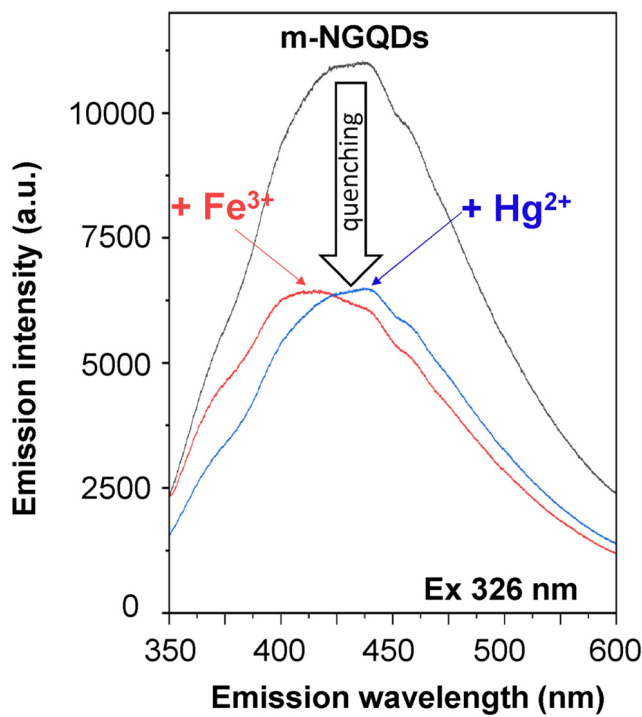

**Fig. 6** Selected emission spectra of the as-functionalized NN-uGQDs obtained with an excitation wavelength of 326 nm in the presence of 0.50 mg L$^{-1}$ of Fe$^{3+}$ and 0.45 mg L$^{-1}$ of Hg$^{2+}$.

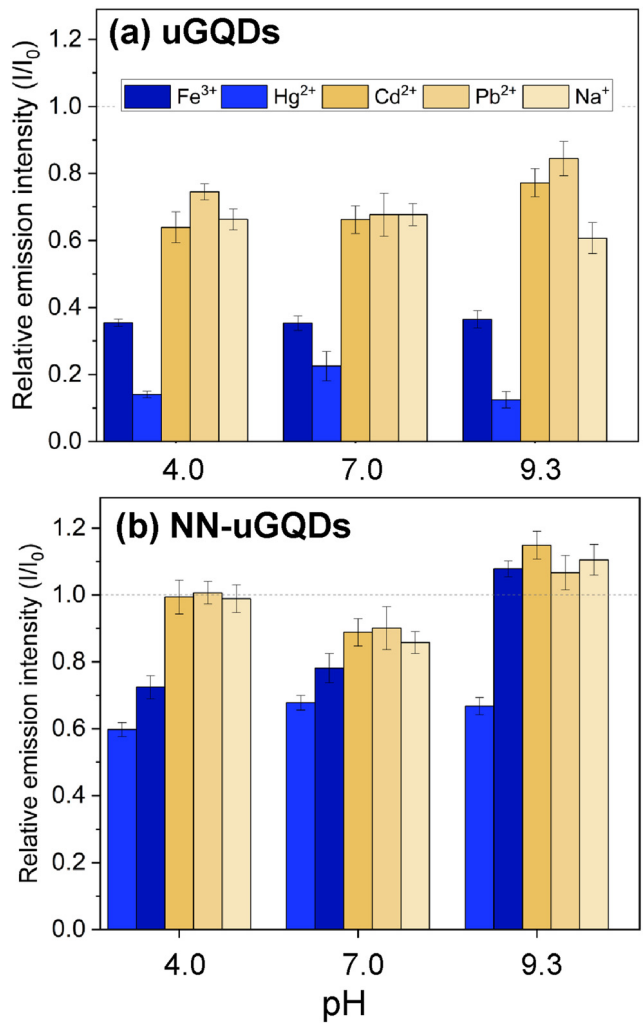

Fig. 7 Relative emission intensities ($I/I_0$) as a function of pH for: a) 0.042 mg mL$^{-1}$ uGQDs and b) 1-nitroso-2-naphthol-modified uGQDs (NN-uGQDs) solutions in the presence of various cations at 1.00 mg mL$^{-1}$ concentration. The maximum emission intensity of pure NN-uGQDs was taken as the reference intensity ($I_0$) and $n$ = 3.

Table 1 shows the selected top-five linear coefficients resulting from the training of the model with the experimental data along with their corresponding $\lambda_{ex}$ and $\lambda_{em}$ pairs. Since these coefficients exhibited the highest values, they can be considered as the most influential for the prediction of $Hg^{2+}$ and $Fe^{3+}$ concentrations. In general terms, these results show that the presence of $Fe^{3+}$ blue-shifted the emission wavelengths, consistent with Fig. 6 and S4 (ESI†).

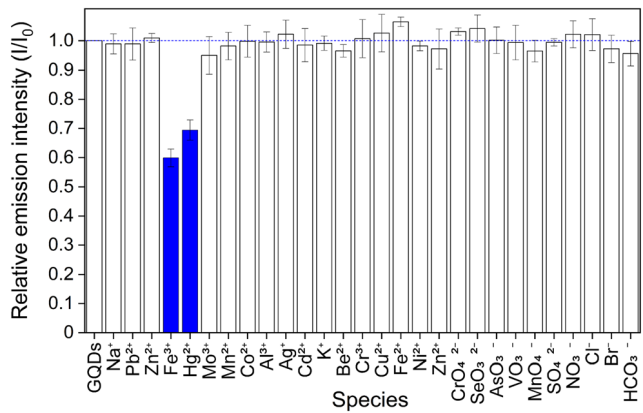

Fig. 8 Relative emission intensity ($I/I_0$) for 0.042 mg mL$^{-1}$ solutions of NN-uGQDs in the presence of 1.00 mg L$^{-1}$ of different added ionic species at pH 4.0. The maximum emission intensity of pure NN-uGQDs was taken as the reference intensity ($I_0$) and $n$ = 3.

### 3.6. Analytical characteristics

The developed ML-based approach required changing the traditional calculation of the analytical figures of merit. For instance, the limit of detection (LOD), which is usually calculated from the calibration curve intercept, now needs to be calculated in a more pragmatic way. This was achieved by measuring triplicates of NN-uGQDs dispersions in the presence of different $Hg^{2+}$ and $Fe^{3+}$ concentrations, and considering the LOD as the lowest concentration needed to quench 3× the STD of the maximum emission intensity of the pure NN-uGQDs, both with (1.0 mg $L^{-1}$) and without the presence of the other analyte. The LODs achieved for $Fe^{3+}$ and $Hg^{2+}$ were 0.003 and 0.001 mg $L^{-1}$ when measured independently and 0.008 and 0.009 mg $L^{-1}$ in the presence of 1.0 mg $mL^{-1}$ of the other species, respectively. Considering a factor of 10 for the calculation of the limit of quantification, the values achieved for $Fe^{3+}$ and $Hg^{2+}$ were 0.01 and 0.003 mg $L^{-1}$ when measured independently and 0.027 and 0.030 mg $L^{-1}$ in the presence of 1.0 mg $mL^{-1}$, respectively.

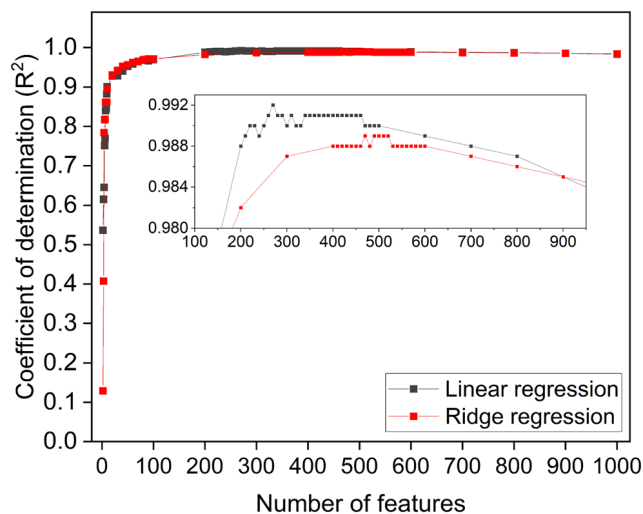

Fig. 9 Average $R^2$ for the cross-validation of linear and ridge regression fitting, as a function of the number of selected features using 191 excitation–emission matrices as data inputs. The hyperparameters alpha and solver were optimized for the ridge model in all iterations. Inset shows $R^2$ values in a reduced range between 0.98 to 1.0 for better comparison between linear and ridge regression.

Table 1 Selected top-5 highest linear coefficients determined using a linear regression model trained with 191 excitation–emission matrices for the prediction of $Hg^{2+}$ and $Fe^{3+}$ concentrations in real samples

| $Hg^{2+}$ | | | $Fe^{3+}$ | | |
|---|---|---|---|---|---|
| Coefficient | $\lambda_{ex}$ (nm) | $\lambda_{em}$ (nm) | Coefficient | $\lambda_{ex}$ (nm) | $\lambda_{em}$ (nm) |
| 0.65 | 342 | 495 | 0.17 | 338 | 403 |
| 0.49 | 350 | 409 | 0.17 | 338 | 483 |
| 0.45 | 314 | 499 | 0.17 | 338 | 449 |
| 0.43 | 310 | 477 | 0.16 | 348 | 435 |
| 0.43 | 312 | 471 | 0.16 | 320 | 427 |

Importantly, when the sum of concentrations from both analytes was above 1.44 mg $L^{-1}$, the accuracy of the model was negatively affected, placing this value at the top limit of the linear range. At this stage, it is important to highlight that the linear range covers most of the expected concentrations of $Fe^{3+}$ and $Hg^{2+}$ in the analyzed samples. For instance, the LODs were well below the 0.3 mg $L^{-1}$ and 0.002 mg $L^{-1}$ for $Fe^{3+}$ and $Hg^{2+}$ in drinking water as recommended by the World Health Organization[50] and the United States Environmental Protection Agency (US EPA),[46] respectively. Moreover, to evaluate the reproducibility of the probe, 6 independent measurements were performed with 0.50 mg $L^{-1}$ of $Fe^{3+}$ and $Hg^{2+}$. This resulted in a 4.2% relative STD, demonstrating very good precision. Measurements performed on five different days, on the other hand, led to a 6.6% STD. Additionally, our results were compared with recently published GQD-based fluorescent sensors for detecting $Hg^{2+}$ and $Fe^{3+}$. Table S1 (ESI†) summarizes information on the fluorescence type of sensor, synthesis method, linear range, LOD, and the source of water employed in other reports. The LODs achieved by our probe exceed most of the values in the table.

### 3.7. Analysis of real samples

To demonstrate its applicability in real life, and to validate the accuracy of the ML model, we quantified the $Hg^{2+}$ and $Fe^{3+}$ present in tap, river, and dam water samples. For this, the linear model was trained with 191 EEMs at known analyte concentrations, and used to predict the unknown concentrations under identical experimental conditions. To validate the results, we employed well-established methods, including CV-AFS and spectrophotometric methods,[51] to precisely measure the $Hg^{2+}$ and $Fe^{3+}$ in real samples. The instrumental conditions for CV-AFS, along with the corresponding external calibration curves used in each method, are detailed in section 7 (Table S2) and (Fig. S6), respectively, as part of the ESI.†

Table 2 shows the results obtained using our ML-based fluorescence approach, and those obtained with spectrophotometric and CV-AFS measurements. This table also features the results of recovery assays carried out for all the samples, including both independent and combined $Hg^{2+}$ and $Fe^{3+}$ standard additions. As can be seen in the table, the results predicted by the ML model are in good agreement with those obtained with the spectrophotometric and CV-AFS methods, clearly justifying the importance of incorporating the ML tool for the simultaneous quantification of the species under study. Furthermore, the results of the recovery assays with the NN-uGQDs probe were satisfactory, with values ranging from 88.0% to 112% for $Fe^{3+}$ and from 94.0% to 108% for $Hg^{2+}$.

Table 2 Simultaneous determination of $Hg^{2+}$ and $Fe^{3+}$ in water taken from tap, river, and dam sources along with recovery studies ($n$ = 3). All the concentrations are in mg $L^{-1}$, unless stated otherwise

| Sample | $Fe^{3+}$ added | $Hg^{2+}$ added | $Fe^{3+}$ (probe + ML) | $Fe^{3+}$ (validation) | $Hg^{2+}$ (probe + ML) | $Hg^{2+}$ (validation, μg $L^{-1}$) |
|---|---|---|---|---|---|---|
| Tap water | 0.00 | 0.00 | 0.371 ± 0.009 | 0.362 ± 0.008 | <LOD | 0.02 ± 0.02 |
| | 0.05 | 0.05 | 0.427 ± 0.008 | 0.416 ± 0.006 | <LOD | 0.02 ± 0.03 |
| | 0.00 | 0.00 | 0.380 ± 0.010 | 0.374 ± 0.009 | 0.049 ± 0.007 | 50.1 ± 0.2 |
| | 0.05 | 0.05 | 0.419 ± 0.009 | 0.410 ± 0.009 | 0.500 ± 0.010 | 509 ± 2 |
| | 0.50 | 0.50 | 0.882 ± 0.007 | 0.888 ± 0.009 | 0.054 ± 0.008 | 49.8 ± 0.3 |
| River water | 0.00 | 0.00 | 0.97 ± 0.01 | 0.96 ± 0.02 | <LOD | 0.05 ± 0.05 |
| | 0.05 | 0.05 | 1.02 ± 0.09 | 1.01 ± 0.09 | <LOD | 0.05 ± 0.04 |
| | 0.00 | 0.00 | 0.97 ± 0.07 | 0.96 ± 0.07 | 0.052 ± 0.009 | 50.4 ± 0.4 |
| | 0.05 | 0.05 | 1.02 ± 0.03 | 1.02 ± 0.02 | 0.510 ± 0.010 | 515 ± 7 |
| | 0.50 | 0.50 | 1.41 ± 0.20 | 1.45 ± 0.20 | 0.047 ± 0.007 | 52.6 ± 0.3 |
| Dam water | 0.00 | 0.00 | 0.771 ± 0.008 | 0.785 ± 0.009 | <LOD | 0.182 ± 0.005 |
| | 0.05 | 0.05 | 0.82 ± 0.01 | 0.833 ± 0.008 | <LOD | 0.188 ± 0.007 |
| | 0.00 | 0.00 | 0.78 ± 0.01 | 0.779 ± 0.009 | 0.054 ± 0.006 | 48.9 ± 0.6 |
| | 0.05 | 0.05 | 0.824 ± 0.008 | 0.84 ± 0.01 | 0.520 ± 0.010 | 496 ± 10 |
| | 0.50 | 0.50 | 1.27 ± 0.05 | 1.27 ± 0.03 | 0.047 ± 0.005 | 521 ± 9 |

## 4. Conclusions

The use of chemically modified GQDs as a fluorescence probe along with the incorporation of ML was hereby developed and applied for boosting the selectivity and simultaneous quantification of $Hg^{2+}$ and $Fe^{3+}$ in natural water samples. GQDs modified with urea were synthesized *via* an electrochemical approach, which yielded highly crystalline nanostructures resembling the CVD graphene precursor. The chemical modification with NN resulted in a remarkable improvement of the selectivity toward $Hg^{2+}$ and $Fe^{3+}$ when compared to the as-synthesized uGQDs, after carefully optimizing the chemical and instrumental parameters. Furthermore, TD-DFT calculations suggested that the NN molecules acted as the fluorophore within the NN-uGQDs, while the $sp^2$ carbogenic core acted as radiation "antennae". Training the ML algorithm with 191 excitation–emission matrices allowed the probe to discriminate between both analytes and to predict their concentrations simultaneously. This process involved the selection of the most relevant $\lambda_{ex}/\lambda_{em}$ combinations and required reducing the dimensionality of the problem to significantly improve the prediction accuracy. Finally, but not least important, the combination of modified GQDs with ML was successfully applied for the simultaneous determination of $Hg^{2+}$ and $Fe^{3+}$ in real samples, including tap, river, and dam waters. Validation measurements *via* a spectrophotometric method and CV-AFS confirmed the accuracy of the ML method. We consider this work as a stepping stone toward the development of new chemical functionalization of GQDs assisted by ML models as an exciting alternative to improve selectivity, sensitivity, and resolution, without the need for tedious experiments or the direct handling of large amounts of instrumental data.

## Author contributions

Mauricio Llaver: conceptualization; data curation; formal analysis; investigation; methodology; visualization; roles/writing – original draft. Santiago D. Barrionuevo: data curation; formal analysis; investigation; visualization; writing – review & editing. Jorge M. Núñez: data curation; formal analysis; investigation. Agostina Chapana: data curation; formal analysis; investigation. Rodolfo G. Wuilloud: funding acquisition; project administration; resources. Myriam H. Aguirre: funding acquisition; project administration; resources; writing – review & editing. Francisco J. Ibañez: conceptualization; funding acquisition; project administration; resources; writing – review & editing.

## Conflicts of interest

There are no conflicts to declare.

## Acknowledgements

This research was funded by Consejo Nacional de Investigaciones Científicas y Técnicas (CONICET) (PIP-2022-2024-0001), Agencia Nacional de Promoción Científica y Tecnológica (FONCYT) (Projects PICT-2019-03859-BID, PICT-2019-2188-BID) and Universidad Nacional de Cuyo (Project 06/M035-T1) and UNLP (Project X-887) (Argentina). We acknowledge Federico Fioravanti and Dr. Gabriela I. Lacconi, from the Instituto de Investigaciones en Fisicoquimica de Córdoba (INFIQC, CONICET/UNC), for their help obtaining and analyzing the XPS spectra. We also acknowledge the financial support of European Commission with the action H2020 RISE projects MELON (Grant No. 872631) and ULTIMATE-I (Grant No. 101007825). Authors would like to acknowledge the use of "Servicio General de Apoyo a la Investigación-SAI, Universidad de Zaragoza".